\documentclass[superscriptaddress,aps,twocolumn]{revtex4-1}
\usepackage{bm}
\usepackage{amssymb}
\usepackage{amsmath}
\usepackage{multirow}
\usepackage{lineno}
\usepackage{graphicx}
\usepackage{soul}
\usepackage[colorlinks=true,citecolor=blue,linkcolor=blue]{hyperref}
\usepackage[usenames]{color}

\begin{document}

\title{Relativistic light synthesis of femtosecond sawtooth pulses}

\author{K. Hu}
\affiliation{Institute for Fusion Theory and Simulation and
Department of Physics, Zhejiang University, Hangzhou 310027, China}
\author{H.-C. Wu}
\thanks{huichunwu@zju.edu.cn}
\affiliation{Institute for Fusion Theory and Simulation and
Department of Physics, Zhejiang University, Hangzhou 310027, China}
\affiliation{IFSA Collaborative Innovation Center, Shanghai Jiao
Tong University, Shanghai 200240, China}

\date{\today}

\begin{abstract}
Formation of femtosecond sawtooth pulses by high harmonic generation
from relativistic oscillating mirrors is studied. For oblique
incidence of $p$-polarized laser pulse, one can efficiently control
the intensity of the first few harmonics by modifying the plasma
density gradient at the plasma-vacuum interface. With appropriate
choice of the laser amplitude and incidence angle, the generated
harmonics can form a millijoule femtosecond sawtooth pulse. The
scheme does not rely on phase manipulation and/or pulse
synchronization as in the state-of-art methods. It is also shown
that such a sawtooth pulse can boost the generation of ultrashort
terahertz pulses in laser-plasma interaction by one order of
magnitude, as compared with that from the conventional two-color
laser pulse.

\end{abstract}
\maketitle

Ultrashort terahertz (THz) electromagnetic pulses can be produced
from the interaction of fs multicolor pulses with gas targets
\cite{Cook,Reimann,Babushkin,Wu1}. Mart\'{i}nez \emph{et al.}
pointed out that by using a fs sawtooth pulse one can increase the
THz radiation efficiency to $2\%$, or 50 times higher than that from
the standard two-color configuration \cite{Martinez}. The sawtooth
pulse can optimize the driven electron trajectories to maximize
their velocities. Realization of fs sawtooth pulses is therefore of
high priority.

A sawtooth wave can be formed by linear superposition of harmonic
waves, and the wave electric field can be written as
\begin{equation}
E(t)=\Sigma^{N}_{k=1}f(t)\frac{E_1}{k}\cos(k\omega_1t-\frac{\pi}{2}),
\end{equation}
where $f(t)$ is the envelope of the sawtooth pulse, and $E_1$ and
$\omega_1$ are amplitude and the frequency of the fundamental
harmonic, respectively, and $k$ is the harmonic order. In fact, to a
high degree of accuracy the sawtooth waveform can be synthesized
with just the fundamental and the next few harmonics. In fact,
Mart\'{i}nez \emph{et al.} \cite{Martinez} found that generation of THz radiation is
most efficient when only the first few harmonics are included, since
the amplitude saturates with addition of higher harmonics. Thus, the
properties of the first few harmonics are of critical importance for
THz radiation wave generation.

Sawtooth lasers can be generated by coherent pulse synthesis
\cite{Hansch,Jiang,Cundiff}, and pulse duration of the order of a
nanosecond and energy of about $1$~mJ has been achieved
\cite{Chan1,Chan2}. Sub-bands with different frequencies are first
obtained from one or several broadband pulses, usually generated by
laser interaction with gas or solid targets, and then undergo phase
manipulation and amplitude manipulation. A successful synthesis
requires manipulation of three parameters: (i) the carrier phase of
each sub-band, (ii) the amplitude of each sub-band, and (iii) the
relative delay between the sub-bands. Currently, two main schemes
are widely adopted \cite{Manzoni}. In the first scheme, each
sub-band undergoes amplitude and phase manipulation in separate
modules before being combined. Since the sub-bands travels over very
different optical paths, relative delay and the phase jitter are
inevitable, making manipulation of the first two parameters quite
challenging. In contrast, in the other approach the sub-bands
propagate and are adjusted as a whole, so that it is difficult to
accurately adjust the amplitudes and the carrier phases.

In this paper we consider generation of intense fs sawtooth laser
pulse by synthesis of relativistically intense laser light through
high-harmonic generation (HHG) from relativistic oscillating mirrors
(ROM). HHG from intense laser pulses interacting with solid targets
has been considered as a promising method for production of bright
ultrashort bursts of x-ray and extreme-ultraviolet radiation
\cite{Teubner,Paul,Dromey,Heissler}. When an ultraintense laser
pulse impinges on a solid target, electrons at the plasma surface
are first driven out into vacuum by the laser field, and then pushed
back towards the target. As the electrons are pulled out of the
plasma, they form an ROM that specularly reflects the driving laser
light, producing a pulse consisting of a series of harmonics
\cite{Linde,Thaury,Lichters,Baeva}. The first few harmonics can have
relativistic intensities.

The major advantage of the proposed scheme is that it avoids
problems associated with phase modulation and pulse synchronization.
The harmonics generated by ROM are locked in phase: there is no
appreciable phase chirping and all harmonics have the same carrier
phase \cite{Quere,Nomura,Kahaly}. Besides, the single laser drive
ensures synchronization and collinearity of the harmonics. The peaks
of the harmonics coincide and the relative delay is almost zero.
That is, synthesis is self accomplished in the process of HHG
generation. The only parameter that needs careful modulation is the
harmonic amplitude, and this can be achieved by introducing an
adjustable exponential density gradient at the target-vacuum
interface \cite{Kahaly}. We have studied the effects of this
gradient length, the target density,  laser intensity, and incidence
angle. And phase properties of the generated harmonics under our
parameters are analysed. Synthesis of several fs relativistic
harmonics makes the sawtooth pulse four to five orders of magnitude
shorter and more powerful than that produced by conventional
coherent pulse synthesis.

HHG from ROM dominates at relativistic intensities, when the
normalized vector potential of the incident laser pulse is close to
or larger than unity \cite{Thaury,Kahaly,Tarasevitch}. According to
the selection rule proposed by Lichters \emph{et al.}
\cite{Lichters}, only oblique incidence of $p$-polarized laser
produces both odd and even harmonics in the same polarization
direction, as required for producing a sawtooth pulse. Our main aim
is to investigate the parameters that are relevant to the harmonic
amplitudes in HHG from ROM. Particle-in-cell (PIC) simulations with
the one-dimensional PIC code JPIC \cite {Wu2}, serve as numerical
experiments for obtaining the harmonic spectra. In particular, we
consider the effects of the (1) laser amplitude
$a_0=eE_0/\omega_0mc$; (2) electron density $n_0/n_c$, where
$n_c=\epsilon_0m\omega_0^2/e^2$ is the critical density; (3) angle
of incidence $\theta$; and (4) density gradient scalelength
$L/\lambda_0$ at the plasma-vacuum interface, where $n\propto
exp(x/L)$. Here $e$ and $m$ are the charge and the mass of the
electron, $c$ is the speed of light. $\lambda_0$ and $\omega_0$ are
the wavelength and the frequency of the incident laser,
respectively.

\begin{figure}[t]
\includegraphics[width=0.3\textwidth]{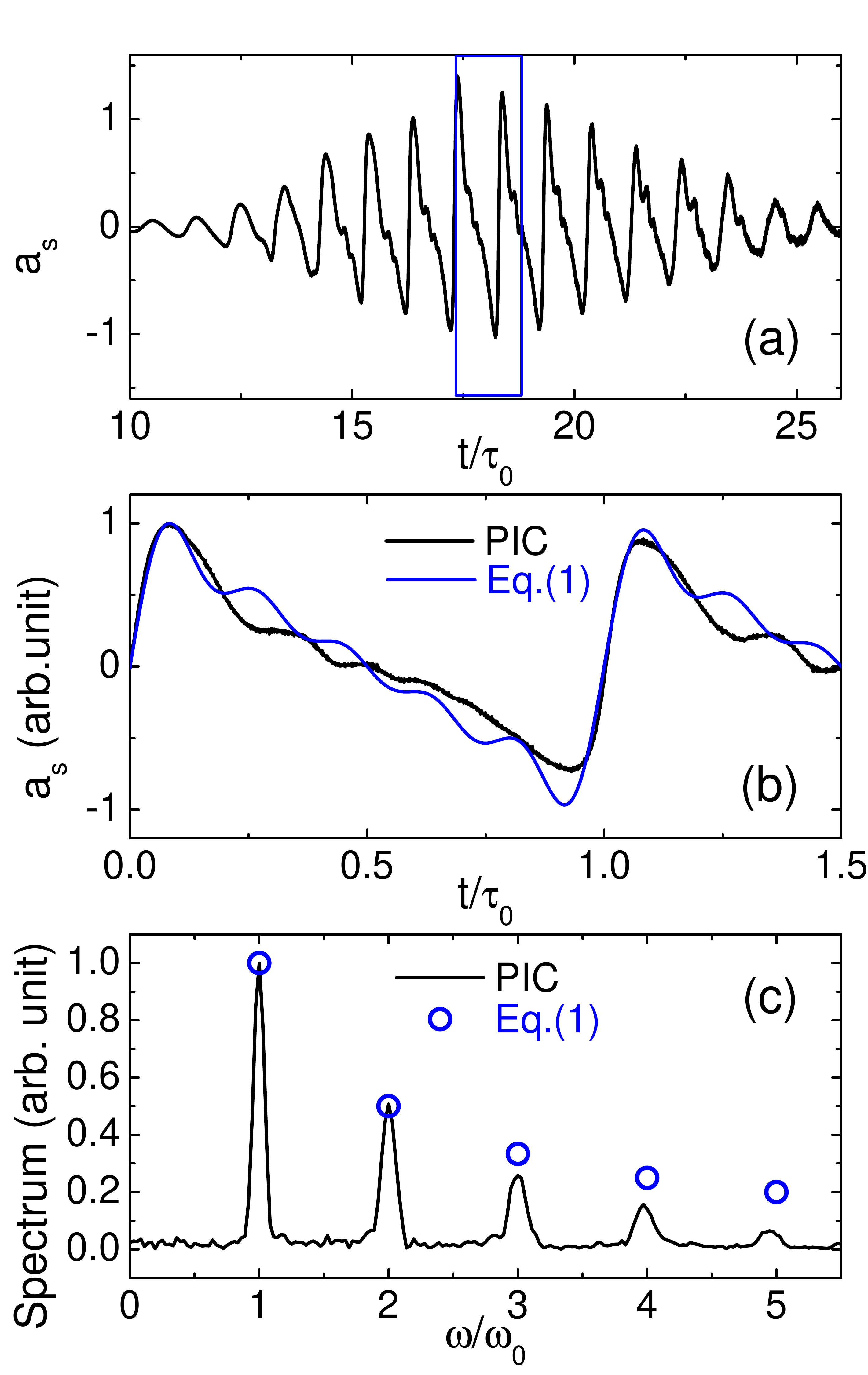}
\caption{(a) Electric field $a_{s}$ of the reflected pulse for
$a_0=1$, $L/\lambda_0=0.6$, $\theta=45^{\circ}$, and $n_0/n_c=100$.
(b) Enlargement of the center part (the blue box in (a)) of the
generated pulse, compared with an ideal sawtooth pulse, obtained
from Eq. (1) with N=5. (c) Spectrum of the pulse in (a) (black
line), compared with the spectrum of an ideal pulse obtained by Eq.
(1) (blue circle). }
\end{figure}

We start with the case where a $p$-polarized $a_0=1$ laser beam
obliquely impinges on a plasma layer at $\theta=45^{\circ}$. The
vector potential of the laser is
$\bm{a}=a_0\cos\lbrack\omega_0(t-x/c)\rbrack exp(-t^2/T^2)\hat{y}$
, where $T=5\lambda_0/c$ is the duration of the
laser. For $\lambda_0=800$ nm, the FWHM intensity of the laser is
$15.75$ fs. The plasma layer is of density $n_0/n_c=100$ and initial
thickness $1.5\lambda_0$. The density gradient scalelength at the
plasma-vacuum surface is $L/\lambda_0=0.6$. The ions remain fixed,
since they move very little on the time scale of interest and thus
have negligible effect on the harmonics.

\begin{figure}[t]
\includegraphics[width=0.5\textwidth]{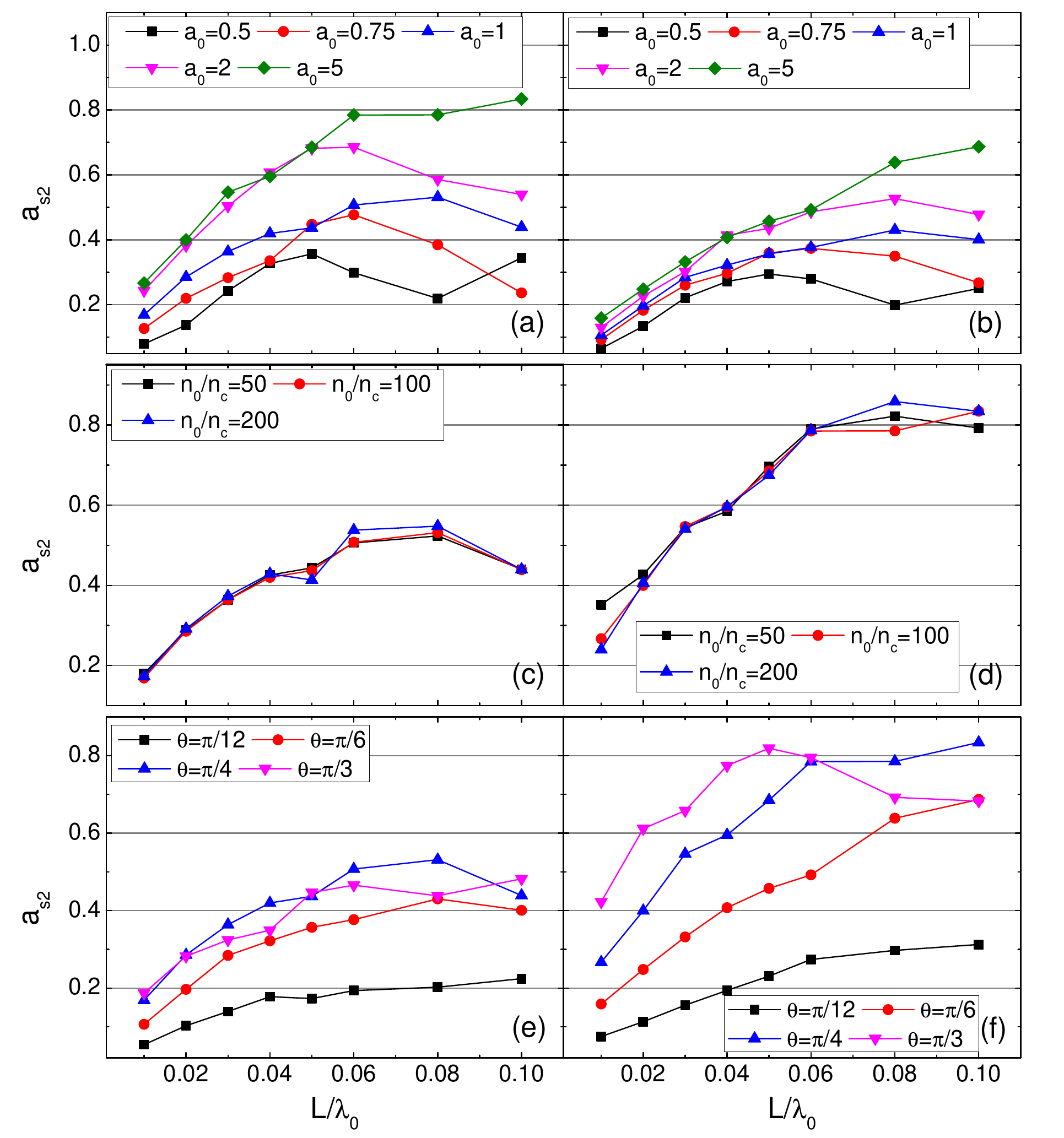}
\caption{Amplitude of the second harmonic of the generated pulse
versus $L/\lambda_0$
for (a) $n_0/n_c=100$, $\theta=45^{\circ}$,
(b) $n_0/n_c=100$, $\theta=30^{\circ}$,
(c) $a_0=1$, $\theta=45^{\circ}$,
(d) $a_0=5$, $\theta=45^{\circ}$,
(e) $a_0=1$, $n_0/n_c=100$, and
(f) $a_0=5$, $n_0/n_c=100$.
}
\end{figure}
The waveform and the spectrum of the reflected light are shown in
Fig. 1. One can see that a nearly Gaussian sawtooth laser pulse is
formed. The pulse has a peak amplitude of $1.4$ (or intensity
$4.2\times10^{18}$~W/cm$^2$). For the source size of $5\lambda_0$,
the pulse has 527 GW in peak power and $9.84$ mJ in energy. The FWHM
of the pulse intensity is $14.9$ fs. From Fig. 1(b), one
can see that the generated waveform is very similar to that obtained
from Eq. (1). This is because the amplitudes of the 2th to 5th
generated harmonics (normalized to the amplitude of the fundamental
harmonic) are $a_{s2}=0.51$, $a_{s3}=0.26$, $a_{s4}=0.16$,
$a_{s5}=0.064$, which are not far from the desired values
$a_{sk}=1/k$, as illustrated in Fig. 1(c). Among them $a_{s2}$
affects the waveform most, thus requiring particular attention. The
harmonics higher than the 4th or 5th have little effect due to their
small amplitudes. In the following, we shall mainly focus on the
second harmonic.

Next we consider the effects of the density scalelength for
$L/\lambda_0=0.01$ to $0.1$, for the laser intensities $a_0=0.5$,
$a_0=0.75$, $a_0=1$, $a_0=2$, and $a_0=5$, with the other parameters
the same as in Fig. 1. For a fixed $a_0$, increasing the scalelength
will first boost up the harmonic amplitude quickly. Then $a_{s2}$
approaches a saturation value and may even decrease a little when
$L/\lambda_0$ gets close to 0.1. The increase can be attributed to
smaller restoring force and larger amplitude of the oscillations
induced by the longer density gradient. (We have excluded the
$L/\lambda_0>0.1$ cases because the corresponding reflected pulses
contain too much noise.) The same trend occurs for the case shown in
Fig. 2(b), where the incidence angle is changed to
$\theta=30^{\circ}$. Both figures also illustrate that emission of
the second harmonic increases greatly for larger laser intensity,
especially for longer scalelength. One can see that $a_{s2}$ falls
off significantly for $a_0=0.5$ and $a_0=0.75$ when
$L/\lambda_0>0.6$. This can be attributed to the fact that coherent
wake emission (CWE) starts to play a role in the process. In order
to produce sawtooth pulses stably, we shall set $a_0 \geqslant 1$.

\begin{figure}[t]
\includegraphics[width=0.5\textwidth]{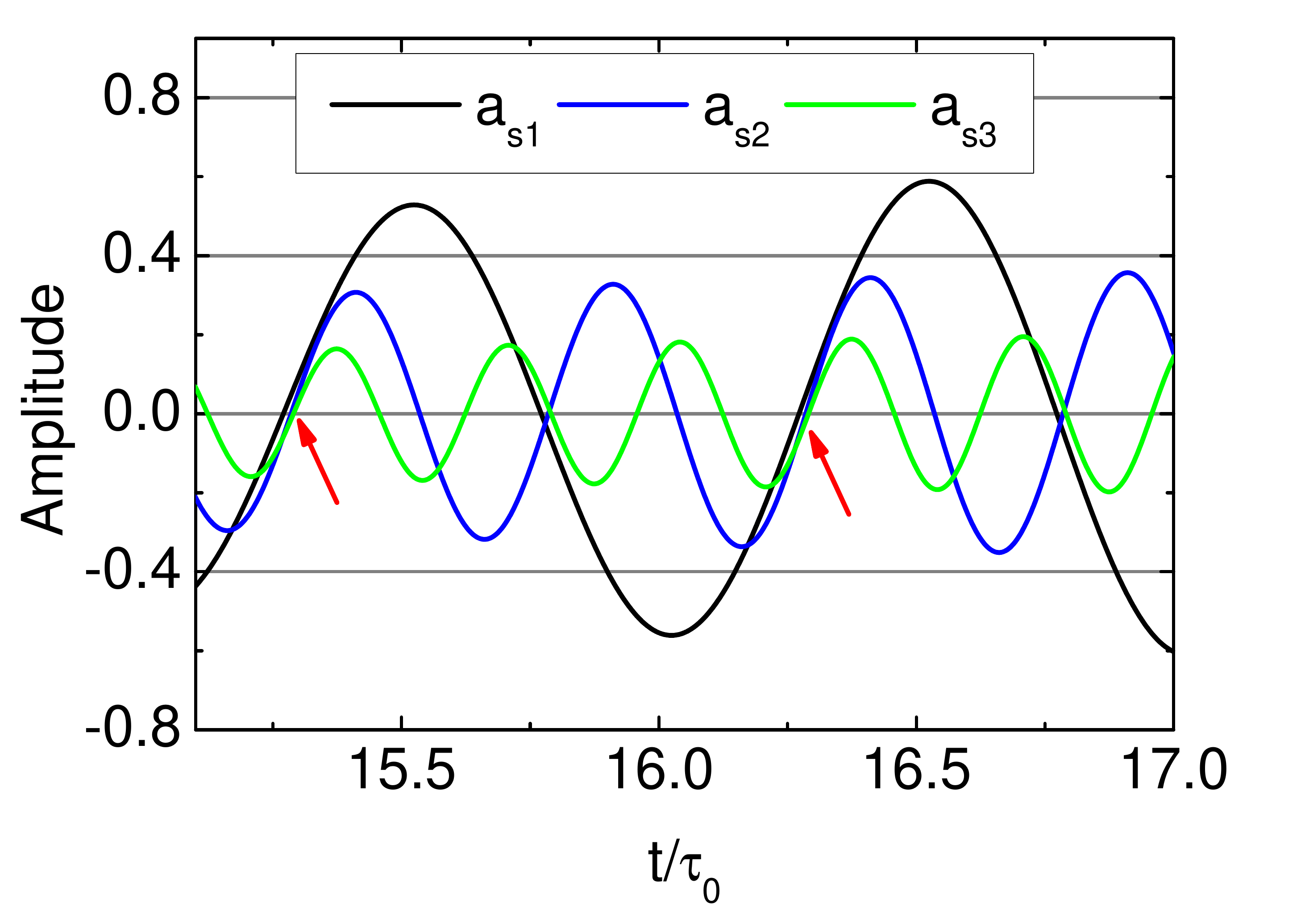}
\caption{
Waveforms of the first (black), the second (blue) and the third (green)
harmonic of the reflected pulse shown in Fig. 2.
}
\end{figure}

To see how the plasma density is related to the harmonic generation,
we have carried out simulations for $n_0/n_c=50, 100,$ and $200$. In
Figs. 2(c) and 2(d), for $a_0=1$ and $5$, respectively, and
$\theta=45^{\circ}$, one can clearly seen that the plasma density
has little effect on the intensity of the second harmonic. This is
because the range of the electron bunch's motion is limited within
the density gradient, thus has little to do with the maximum density
of the plasma target. Figures 2(e) and 2(f) show the relationship
between the incidence angle and the amplitude of the second
harmonic. Electrons of the ROM are driven by the ponderomotive
force, determined by the vertical component of the laser field, as
well as by direct action of the parallel component. PIC simulations
show that the total contribution of these two effects boosts the
harmonic amplitude to a maximum at $\theta=45^{\circ}$ to
$60^{\circ}$.

Another point that should be considered is the relative phases of
the harmonics. Previous studies have shown that harmonics have no
phase difference in the case $L/\lambda_0 \ll0.1$ \cite{Quere}. To
test if this conclusion still stands when a longer gradient of
$L/\lambda_0\sim0.1$ is introduced, we have analyzed the phases of
the first three harmonics in the case shown in Fig. 1(a). The
results are visualized in Fig. 3. Each curve represents the waveform
of one harmonic, obtained by Fourier transform, containing
information of the frequency, amplitude, and phase. One can see that
the phase difference of the harmonics is almost zero (marked by the
red arrows), in agreement with Eq. (1). That is, in our scheme the
phase properties of the harmonics naturally meet the condition for a
sawtooth wave.

\begin{figure}[t]
\includegraphics[width=0.46\textwidth]{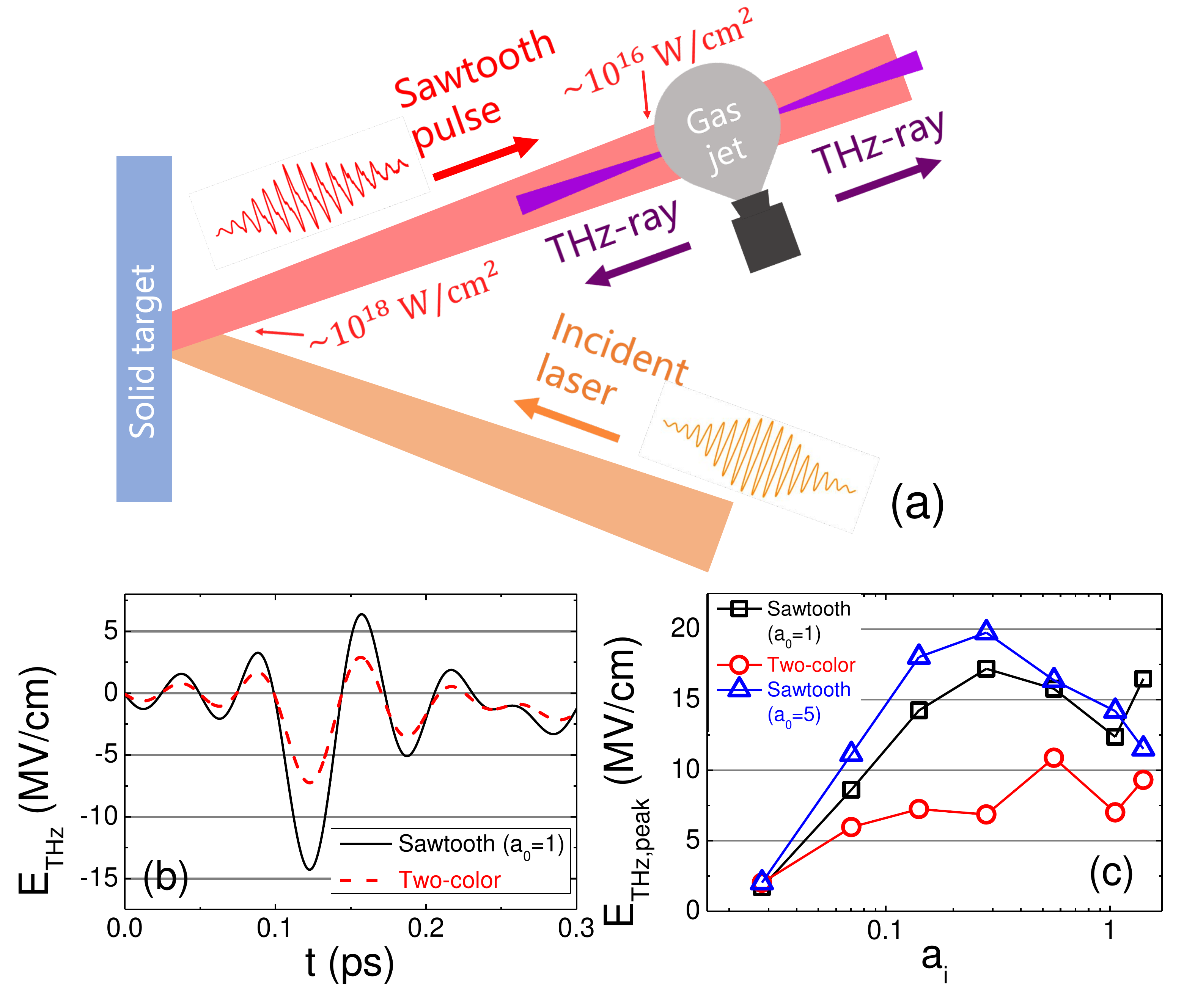}
\caption{ (a) Schematic setup of THz radiation generation by sawtooth
pulses. (b) The THz radiation pulses from the sawtooth and two-color
cases for $a_i=0.14$. (c) Peak electric fields of the generated THz
radiation for different laser intensities. }
\end{figure}

As discussed above, fs sawtooth pulses can be produced by choosing
appropriate parameters such as the density gradient scalelength,
laser intensity and incidence angle. As the density gradient
scalelength becomes longer, PIC simulations first show a fast rise
of harmonic intensities, follow by progressive saturation, when the
scalelength is larger then $0.06\lambda_0$. The saturation value
increases for larger laser intensity and reaches its maximum for an
incidence angle between $45^{\circ}$ to $60^{\circ}$. The key point
is to adjust the saturation value of the second harmonic to near 0.5
by selecting proper laser amplitude and incidence angle. No phase
manipulation and synchronization modules are needed and the
generated sawtooth pulse can be directly impinged on the gas jet for
THz radiation production. The schematic drawing is given in Fig.
4(a).

Next we check if the obtained sawtooth pulse can boost THz radiation
generation from a He target, using 1D PIC simulation. The process of
ionization is added to the code and the ionization potentials are
$24.587$ eV for He$\to$He$^+$ and 54.416\ eV for
He$^+$$\to$He$^{2+}$. Two sets of simulations are performed, and
each set contains seven simulations with different laser
intensities. In the first set, the sawtooth wave shown in Fig. 1(a)
is used as the incident pulse, and the field strengths are
$a_{i1}=1.4$, $a_{i2}=0.6a_{i1}=0.84$, $a_{i3}=0.4a_{i1}=0.56$,
$a_{i4}=0.2a_{i1}=0.28$, $a_{i5}=0.1a_{i1}=0.14$,
$a_{i6}=0.05a_{i1}=0.07$ and $a_{i7}=0.02a_{i1}=0.03$. In the
experiments, one can adjust the laser strength by changing the
distance between the laser's focusing point and the target. For
comparison, in the second set we use Gaussian $p$-polarized,
two-color laser pulses with the temporal profile $a_1\times
exp(-t^2/T_1^2)cos[\omega_1[t-x/c)-\pi/2]+a_1/2\times
exp(-t^2/T_1^2)cos[2\omega_1(t-x/c)-\pi/2]$. We consider seven
values for $a_1$, and set $\lambda_1=2\pi c/\omega_1=800nm$ and
$T_1=2.74\lambda_1/c$ to ensure that these pulses have the same
powers and the same maximum field strengths as the corresponding
ones in the first set. The simulation box is 70$\lambda_1$ in the
$x$-direction and the cell number per $\lambda_1$ is 100. The He
target has a length of $10\lambda_1$ and its atomic density is
0.00125$n_c$. After full ionization, the plasma density is
$n_e=0.0025n_c$ and the corresponding plasma frequency is $18.7$
THz. We concentrate on the backward THz radiation with
frequency lower than 20 THz. The higher frequency radiation is
blocked by a low-pass filter.

Figure 4(a) shows the THz radiation for two types of incident
lasers. The lasers have a peak amplitude of $a_i=0.14$,
corresponding to an intensity of $I_i=4.21 \times 10^{16}$ W/cm$^2$.
For such high laser strength, $99.5\%$ of the electrons contained in
neutral He atoms are ionized by the incident laser pulse. It is
obvious that the THz radiation in the sawtooth case has a peak
electric field of $14.24$ MV/cm, 2 times larger than that in the
two-color case. Figure 4(b) summarizes the peak amplitudes of the
THz radiation for different laser intensities in both cases. We see
that with $a_i>0.7$ ($\sim I_i>10^{16}$ W/cm$^2$), the sawtooth
pulse obtained from our HHG scheme can effectively increase the THz
radiation field strength by 2 to 3 times, compared to that from the
two-color pulse. If the spot sizes are the same in both cases, the
efficiency boost can be several times to one order of magnitude
larger. The boost cannot be observed when $a_i=0.03$, since the
laser is too weak and less than $6\%$ of the target electrons are
ionized.

We have also carried out a third set of simulations, in which the
approximate sawtooth wave generated by the $a_0=5$ laser pulse (the
green diamonds in Fig. 2(a)) is adopted. Again, the maximum electric
fields are the same as in the previous two sets. The results are
shown by the blue triangles in Fig. 4(c). Surprisingly, the
resulting THz radiation pulses have even larger amplitudes than that
in the first set. This can be attributed to the fact that the
incident pulses in the third set contains stronger higher-order
harmonics as compared to the sawtooth pulses used in the first set.

In summary, in this paper we have presented a robust method for
generating a sawtooth pulse via synthesis of relativistic fs
harmonics. The latter are generated by laser interaction with plasma
targets. By introducing a density gradient scalelength of
$0.06\lambda_0-0.1\lambda_0$ at the plasma-vacuum interface, we can
effectively increase the amplitudes of the first few harmonics to
meet the requirement for forming a sawtooth wave. By proper choice
of laser amplitude and incidence angle, a fs gigawatt sawtooth pulse
can be generated. Such a sawtooth pulse can boost the generated THz
radiation by several times, compared with that from the two-color
pulse with the same power. Finally, we also note that the present
scheme completely avoids the complicated process of phase modulation
and pulse synchronization that is required in the conventional
methods for generating THz radiation.

\begin{acknowledgments}
This work was supported by the Thousand Youth Talents Plan, NSFC (No.
11374262, 61627901), and Fundamental Research Funds for the Central
Universities.
\end{acknowledgments}

\end{document}